\documentclass[a4paper,12pt]{article}
\usepackage{epsfig}
\usepackage{psfig}
\date{\today}
\newcommand{\bmat}{\left(\begin{array}}
\newcommand{\emat}{\end{array}\right)}
\newcommand{\be}{\begin{equation}}
\newcommand{\ee}{\end{equation}}
\newcommand{\bea}{\begin{eqnarray}}
\newcommand{\eea}{\end{eqnarray}}

\def\lsim{\raise0.3ex\hbox{$\;<$\kern-0.75em\raise-1.1ex\hbox{$\sim\;$}}}
\def\gsim{\raise0.3ex\hbox{$\;>$\kern-0.75em\raise-1.1ex\hbox{$\sim\;$}}}
\def\Frac#1#2{\frac{\displaystyle{#1}}{\displaystyle{#2}}}

\begin{document}
\pagestyle{empty}
\renewcommand{\thefootnote}{\fnsymbol{footnote}}
%\begin{titlepage}
%\pagestyle{empty}
\rightline{CERN-TH/2002--008}  
\rightline{DCPT/02/12}
\rightline{FTUAM 02/02}
\rightline{IFUM-76/FT} 
\rightline{IFT-UAM/CSIC-02-03}
\rightline{IPPP/02/06} 
%\rightline{MILANO-02-??}
\rightline{February 2002}
%\vskip 1cm

\vspace{.3cm} {\Large
\begin{center}
{\bf Relic Neutralino Density in Scenarios with Intermediate Unification Scale}
\end{center}}
\vspace{.3cm}

\begin{center}
S. KHALIL~$^{1,2}$\ , C. MU\~NOZ~$^{3,4,5}$ and E. TORRENTE-LUJAN~$^{3,5,6}$ \\
\vspace{.7cm}

1. \emph{IPPP, Physics Department, Durham University, DH1 3LE,
Durham, U.K.}
%\\
\vspace{.2cm}

2. \emph{Ain Shams University, Faculty of Science, Cairo, 11566,
Egypt.}

\vspace{.3cm}

3. \emph{Departamento de F\'{\i}sica
Te\'orica C-XI, Universidad Aut\'onoma de Madrid,\\
Cantoblanco, 28049 Madrid, Spain.}

\vspace{.3cm}

4. \emph{Instituto de F\'{\i}sica Te\'orica  C-XVI,
Universidad Aut\'onoma de Madrid,\\
Cantoblanco, 28049 Madrid, Spain.}

\vspace{.3cm}

5. \emph{Theory Division, CERN, 1211 Geneva 23, Switzerland.}

\vspace{.3cm}

6. \emph{Dipartimento di Fisica,
Universita degli Studi di Milano,\\
Via Celoria 16, Milano, Italy.}

\end{center}

\vspace{.7cm}
%\hrule \vskip 0.3cm
\begin{center}
\small{\bf Abstract}\\[3mm]
\end{center}

We analyse the relic neutralino density in supersymmetric models with
an intermediate unification scale. In particular, 
we present concrete cosmological scenarios where the reheating
temperature is as small as $\cal{O}$($1 - 1000$ MeV).
%1 MeV -- 1 GeV. 
% ($M_I\sim 10^{12}$) .
When this temperature is associated to 
the decay of moduli fields producing neutralinos,
%appearing in string theories, 
we show that the relic abundance increases considerably 
with respect to 
the standard thermal production. 
Thus the neutralino becomes a good dark matter
candidate with $0.1\lsim \Omega h^2 \lsim 0.3$, 
even for regions of the parameter space where
large neutralino-nucleon cross sections, compatible with
current dark matter experiments, are present. 
This is obtained
for intermediate scales $M_I\sim 10^{11}-10^{14}$ GeV, and 
moduli masses $m_\phi\sim 100-1000$ GeV.
On the other hand, when the above temperature is associated to
the decay of an inflaton field, the relic abundance is
too small.

\begin{minipage}[h]{14.0cm}
\end{minipage}
%\vskip 0.3cm \hrule \vskip 1cm

\newpage
%----------------------------------------------------------------------%
%  Resetting of counters
%----------------------------------------------------------------------%
\setcounter{page}{1}
\pagestyle{plain}
\renewcommand{\thefootnote}{\arabic{footnote}}
\setcounter{footnote}{0}
%----------------------------------------------------------------------%

\section{Introduction}

% {\bf 1.}
As it is well known, the lightest neutralino, $\tilde\chi_1^0$,
is a weakly interacting massive particle (WIMP), and therefore
a very interesting candidate for dark matter in the 
universe. In fact, many experimental efforts are being
carried out in order to detect WIMPs through elastic scattering with
nuclei in a detector \cite{contemporary}.
In this sense the theoretical analysis of the neutralino--nucleus
cross section 
$\sigma_{\tilde\chi_1^0-N}$ 
is very important. In particular, these analyses
in the context of the minimal
supersymmetric standard model 
(MSSM) are usually
performed assuming the unification scale $M_{GUT} \approx 10^{16}$
GeV for the running of the universal soft supersymmetry (SUSY)--breaking
terms. 
However,
%In some recent works \cite{interm1,interm2}, various
%implications for an  intermediate unification scale in
%supersymmetric models have been considered. In particular, 
it was
pointed out recently \cite{interm1} that this cross section 
is very sensitive to the variation of the unification scale. For
instance, by taking an intermediate unification scale
$M_I \approx 10^{10-12}$ GeV the cross section increases substantially,
being compatible for large regions of the parameter space of the MSSM
with the sensitivity of current dark matter experiments 
$\sigma_{\tilde\chi_1^0-N}\approx 10^{-7}$--$10^{-6}$ GeV$^{-2}$, 
for $\tan\beta\gsim 3$ and 
$m_{\tilde\chi_1^0}\approx 100$ GeV.
For larger values of the scale, as e.g. $M_I=10^{14}$ GeV,
a similar result is obtained for $\tan\beta\gsim 10$.
Explicit scenarios with intermediate scales, 
arising in D-brane constructions from type I strings, were analysed 
in ref.~\cite{interm2}.
Although compatibility with the experiments 
may also be obtained within the usual MSSM 
scenario with
the scale $\approx 10^{16}$ GeV, it requires large values of $\tan\beta$ 
($\tan \beta \gsim 20$) \cite{Bottino}-\cite{Mario}
or a specific non--universal structure of the 
soft terms \cite{Bottino,Arnowitt,Nath2,darkcairo}.

%are
%enhanced and a large region of the parameter space of the minimal
%supersymmetric standard model (MSSM) is compatible with the
%sensitivity of the current dark matter detectors. In fact, such
%results can also be obtained within the GUT unification scenario,
%\ie, $M_I \sim 10^{16}$ GeV, but it requires either large $\tan
%\beta$ ($\tan \beta \ge 25$) or specific choice for non--universal
%soft SUSY breaking terms so that the value of $\mu$ term is
%reduced and hence a larger Higgsino components of the LSP is
%obtained, which is essential for enhancing the LSP--nucleon cross
%sections \cite{interm1}.

In all the above works the relic neutralino density was also discussed.
%%Unlike the case of the MSSM with the 
%GUT scale, universality, and moderate values
%of $\tan\beta$, were there is always a set of 
%parameters which yield 
%the relic density 
%within the observational bounds
%$0.1\lsim \Omega_{\tilde\chi_1^0}h^2\lsim 0.3$, in these scenarios
%with a large cross section in some regions of the parameter space,
%generically
%$\Omega_{\tilde\chi_1^0}h^2\lsim0.01$.
In these scenarios
with a large cross section in some regions of the parameter space,
generically
$\Omega_{\tilde\chi_1^0}h^2\lsim0.01$.
Of course, this might be a potential problem for the consistency of
those
regions 
given the observational bounds\footnote{It 
is worth noticing, however, that more conservative
lower bounds, $\Omega_{\tilde\chi_1^0}h^2\approx 0.01$,
have also been quoted in the literature. For a brief discussion
on this issue see e.g. ref.~\cite{darkcairo} and references therein.}
$0.1\lsim \Omega_{\tilde\chi_1^0}h^2\lsim 0.3$.

This result is obtained because in the usual early--universe model
thermal production of neutralinos gives rise to
$\Omega_{\tilde\chi_1^0} h^2 \propto  
1/\langle \sigma^{\mathrm{ann}}_{\tilde\chi_1^0} v \rangle$,
where $\sigma^{\mathrm{ann}}_{\tilde\chi_1^0}$ is 
the cross section for annihilation of a pair of neutralinos, 
$v$ is the relative velocity between the two neutralinos, 
and $\langle .. \rangle$ denotes thermal averaging.
Therefore, in this scheme the relic density
is inversely proportional to the annihilation cross section.
% one usually obtains 
%\be \Omega_{\chi} h^2
%\simeq \Frac{C}{\langle \sigma^{\mathrm{ann}}_{\chi} v \rangle }~,
%\ee 
Let us recall that crossing arguments,
when the main annihilation channel is into quarks, 
ensure that the cross sections of 
annihilation and scattering with nucleons are similar. Thus a large
scattering cross section $\sigma_{\tilde\chi_1^0-N}$
leads generically to a large
annihilation cross section $\sigma^{\mathrm{ann}}_{\tilde\chi_1^0}$,
and as a consequence to a small relic density.

%as Eq. 1 suggest.
%Indeed, in these models, it was observed that LSP--proton cross
%sections of order $10^{-6}$ Pb correspond to low relic densities
%($<\sim 0.005-0.01$ ). For the sake of consistency, 
% one would have  to assume that not all the
%dark matter in our Galaxy are neutralinos and other candidates are
%needed.

However, it is important to remark that this result 
depends on assumptions about the evolution of the early 
universe. In principle, different cosmological scenarios might give rise to
different results. To address this question is precisely
the aim of this paper. We will study the relic density in the 
context of some non-standard 
cosmological scenarios. 
In particular, we will show that, when
intermediate scales are present, results
different from the usual ones summarised above may be produced.
This is because a low reheating temperature, below
the freeze-out temperature, can be obtained.
We will see that, in the case of one of the scenarios, 
values of the relic density within the observational bounds
are possible, even for regions of the parameter space with
a large neutralino--nucleus cross section
$\sigma_{\tilde\chi_1^0-N}\approx 10^{-7}$ GeV$^{-2}$.

The content of the paper is as follows.
In Section~2 we will briefly review the usual cosmological scenario
where thermal production of neutralinos is assumed.
Several well-known formulas will be explicitly written since
we will use them in the discussions of the next sections. 
Then, in Section~3,
we will discuss the modifications introduced in the relic density
analysis
by considering non--standard cosmological scenarios in the case of
intermediate scales.
In particular, we will study the situation when an inflation or a
modulus field
produce low reheating temperatures, close to the nucleosynthesis one.
Finally, the conclusions are left for Section~4.

\section{Thermal production of neutralinos}

Let us briefly review the standard computation of the cosmological
abundance of neutralinos \cite{report}.
Neutralinos were in thermal equilibrium with the standard
model particles in the early universe, and decoupled when
they were non-relativistic.
The process was the following.
When the temperature $T$ of the universe was larger than the mass of the
neutralino, 
the number density of neutralinos and photons was roughly the same, 
$n_{\tilde\chi_1^0}^{eq}\propto T^3$,
and
the neutralino was annihilating with its own antiparticle into lighter
particles and vice versa.
However, shortly after the temperature dropped below the mass of the
neutralino,
$m_{\tilde\chi_1^0}$,
its number density dropped exponentially,
$n_{\tilde\chi_1^0}^{eq}\propto e^{-m_{\tilde\chi_1^0}
%_{\mbox{\tiny WIMP}}
/T}$,
because only a small fraction of the light particles mentioned above
had sufficient kinetic energy to create neutralinos.
As a consequence, the neutralino annihilation rate 
$\Gamma_{\tilde\chi_1^0} = 
\langle\sigma^{ann}_{\tilde\chi_1^0} v \rangle n_{\tilde\chi_1^0}$
dropped below the expansion
rate of the universe, $\Gamma_{\tilde\chi_1^0} \lsim H$, where
$H$ is the Hubble expansion rate. At this point  
neutralinos came away, they could not annihilate, and
their density is the same since then. This can be obtained 
using the Boltzmann equation:
\be 
\frac{d n_{\tilde\chi_1^0}}{d t} + 3 H n_{\tilde\chi_1^0} = - \langle
\sigma^{ann}_{\tilde\chi_1^0} v \rangle \left[ (n_{\tilde\chi_1^0})^2
-(n_{\tilde\chi_1^0}^{eq})^2 \right]\ . 
\label{boltzmann} 
\ee 
One can discuss qualitatively the solution using
the freeze-out condition
$\Gamma_{\tilde\chi_1^0} = 
{\langle\sigma^{ann}_{\tilde\chi_1^0} v \rangle}_F n_{\tilde\chi_1^0}=H$.
Then 
$\Omega_{\tilde\chi_1^0} h^2 =(\rho_{\tilde\chi_1^0}/\rho_{c}) h^2$, 
where $\rho_{\tilde\chi_1^0}$ is the current neutralino mass density and 
$\rho_{c}$ is the critical density, 
turns out to be
\begin{eqnarray}
\Omega_{\tilde\chi_1^0}
%^{(ann)} 
h^2  =
\frac{m_{\tilde\chi_1^0}\ H}
{(2 \pi^2/45)\ g_\star(T_F)\ T_{F}^3\  
% {
\langle \sigma^{ann}_{\tilde\chi_1^0} v  \rangle_F}
% }
\ 
\frac{h^2}{ \rho_c /s_0}=
\frac{m_{\tilde\chi_1^0}\ (45/6\pi \sqrt{10})}
{M_P\ g_\star^{1/2}(T_F)\ T_{F}\  
% {
\langle \sigma^{ann}_{\tilde\chi_1^0} v  \rangle
% }
_F}\ 
\frac{h^2}{ \rho_c /s_0}
\ ,
\label{thermal}
\end{eqnarray}
where 
$g_*(T)$ is the effective number of
degrees of freedom at temperature $T$
(e.g. including all the standard-model degrees of freedom
one gets $g_*=106.75$), 
$T_F$ is the freeze-out temperature, and
$s_0$ is the current entropy density. 
In the second expression we have used the fact that
%$H=\left(\frac{\pi^2 g_*(T)}{90}\right)^{1/2}/M_P$
$H=\left(\pi^2 g_*(T)/90\right)^{1/2} T^2 M_P^{-1}$, with 
$M_P=M_{Planck}/\sqrt{8\pi}\simeq 2.4\times 10^{18}$ GeV the reduced Planck mass.
Taking into account the current value
$\rho_c/s_0\simeq 3.6\times 10^{-9}$ GeV$\times h^2$, 
and the typical freeze-out temperature $T_F\simeq m_{\tilde\chi_1^0}/20$,
one can write the above expression as
\bea
\Omega_{\tilde\chi_1^0}
%^{(ann)} 
h^2 \simeq 1.7 \times 10^{-10}
\left(\frac{1\ GeV^{-2}}
{\langle \sigma^{ann}_{\tilde\chi_1^0} v  \rangle}\right)
\left(\frac{100}{g_*(T_{F})}\right)^{1/2} 
\ . 
\label{thermal2} 
\eea 
Since neutralinos freeze--out at 
$T_F\simeq m_{\tilde\chi_1^0}/20<<m_{\tilde\chi_1^0}$, 
they are non--relativistic and therefore the 
averaged annihilation cross section can be expanded as
%$\langle \sigma^{ann}_{\chi} v \rangle$ 
follows: 
\be 
\langle \sigma^{ann}_{\tilde\chi_1^0} v \rangle = \alpha_s +\alpha_p 
\langle v^2\rangle\ , 
\label{sigma} 
\ee 
where $\alpha_s$ describes the 
s-wave annihilation
and $\alpha_p$ describes both s-- and p-- wave annihilation.
Then, eq. (\ref{thermal2}) can alternatively be 
written as:
\bea
\Omega_{\tilde\chi_1^0} h^2  & \simeq & 
%\Frac{2.8 \times 10^8\ 
% (m_{\tilde\chi_1^0}/GeV)}
% {0.264\sqrt{8\pi}\ g_*^{1/2}(T_F) m_{\tilde\chi_1^0} M_P \left(\alpha_s/x_F + 3
%\alpha_p/x_F^2\right)}
%\nonumber\\
%& = &
8.8 \times 10^{-11}
\Frac{GeV^{-2}}
{g_*^{1/2}(T_F) \left(\alpha_s/x_F + 3
\alpha_p/x_F^2\right)}
\ , 
\label{omega1} 
\eea
where $x_F \equiv  m_{\chi}/T_F$.

As it is well known,
in most of the parameter space 
of the MSSM the neutralino is mainly pure bino, and
as a consequence it will mainly annihilate into
lepton pairs through $t$--channel exchange  of right--handed
sleptons. The $p$--wave dominant cross section is given 
by \cite{olive,kolb}
\bea
\langle \sigma^{ann}_{\tilde\chi_1^0} v  \rangle 
\simeq 8\pi\alpha'^2
%\frac{g'^4}{2\pi}
\frac{1}{m_{\tilde\chi_1^0}^2}
\frac{1}{\left(1+x_{\tilde l_R}\right)^2}
\ \langle v^2  \rangle 
\ , 
\label{annihil} 
\eea 
where $x_{\tilde l_R}\equiv m_{\tilde l_R}^2/m_{\tilde\chi_1^0}^2$
and $\alpha'$ is the coupling constant for the $U(1)_Y$ interaction.
Taking $m_{\tilde l_R}\sim m_{\tilde\chi_1^0}\sim 100$ GeV,
$\langle \sigma^{ann}_{\tilde\chi_1^0} v  \rangle$ in 
eq.~(\ref{annihil}) becomes of the order of $10^{-9}$ GeV$^{-2}$ or smaller.
Using eq.~(\ref{thermal2}) an interesting relic abundance, 
$\Omega_{\tilde\chi_1^0}h^2\gsim 0.1$, 
is obtained.

However, in the special regions mentioned in the Introduction,
with non-universality and/or large $\tan\beta$, 
the lightest neutralino may have an important Higgsino component,
producing a larger cross section. 
This is also the case of scenarios with an intermediate unification
scale. The upper bound for the annihilation cross section,
obtained when the neutralino is Higgsino--like, 
is given by \cite{olive}
\bea
\langle \sigma^{ann}_{\tilde\chi_1^0} v  \rangle 
%\simeq \frac{g_2^4}{2\pi}
\simeq \frac{\pi\alpha_2^2}{2}
\frac{1}{m_{\tilde\chi_1^0}^2}
\frac{\left(1-x_{W}\right)^{3/2}}{\left(2-x_{W}\right)^2}
\ , 
\label{annihil2} 
\eea 
where $x_{W}\equiv m_{W}^2/m_{\tilde\chi_1^0}^2$
and $\alpha_2$ is the coupling constant for the $SU(2)_L$ interaction.
Here one is considering that the Higgsino dominantly annihilates into 
W-boson pairs.
Since now 
$\langle \sigma^{ann}_{\tilde\chi_1^0} v  \rangle$
in eq.~(\ref{annihil2}) is of the order of $10^{-8}$ GeV$^{-2}$,
the relic abundance given by eq.~(\ref{thermal2}) turns out to be small,
$\Omega_{\tilde\chi_1^0}h^2\approx 0.01$, as expected.

\section{Non--standard cosmological scenarios}

In the standard computation reviewed in the previous section, one is tacitly
assuming that the radiation--dominated era is the result of a reheat 
process in the early universe, where the reheating temperature
$T_{RH}$ is very large, in particular
$T_{RH}>>T_F\sim 10$ GeV. The scalar field $\phi$, whose decay leads to
reheating, is usually assumed to be the inflaton field.
One can estimate the reheating temperature as a function
of the decay width $\Gamma_\phi$ as \cite{early}
\be 
T_{RH}= \left(\frac{90}{\pi^2 g_*(T_{RH})}\right)^{1/4} 
(\Gamma_\phi M_P)^{1/2}\ .
\label{trh}
\ee 

However, the only constraint on the reheating temperature is
$T_{RH}\gsim 1$ MeV in order not to affect the successful 
predictions of big--bang nucleosynthesis. This allows in principle
to consider cosmological scenarios with a low reheating 
temperature \cite{kolb}, $T_{RH}<T_F$.
On the other hand, the reheating process can also be associated
with the decay of moduli fields, as e.g. those appearing in string
theory.
Thus the relic abundance could receive contributions from this 
source \cite{moroi}.
 
In what follows, we will show that scenarios with intermediate
unification scales are explicit examples
for the two non--standard cosmological possibilities mentioned above,
namely, decay of the 
inflaton and moduli fields producing a low reheating temperature.
For this analysis eq. (\ref{trh}) is still valid using $\Gamma_\phi$
as given by the corresponding scenario.
This is also true for relation
$\Omega_{\tilde\chi_1^0} h^2 \propto  
1/\langle \sigma^{\mathrm{ann}}_{\tilde\chi_1^0} v \rangle$ in the
case of neutralino production through modulus decay.
Notice that in this case
$\Omega_{\tilde\chi_1^0}\propto 1/T_F$, as shown in eq. (\ref{thermal}),
and therefore if this mechanism produces
a temperature smaller than the typical 
$T_F\simeq m_{\tilde\chi_1^0}/20$, the value of the
relic neutralino density will be increased 
\footnote{For another alternative cosmological scenario with the
potential of increasing the relic density see
ref.~\cite{Brandenberger},
where the decay of cosmic strings producing neutralinos is considered.
}.
We will see below that temperatures as required can be obtained
in scenarios with intermediate scales.
In ref.~\cite{moroi} 
this mechanism was applied in order to obtain reasonable values of 
the relic wino density in
anomaly-mediated SUSY breaking scenarios, using 
$m_{\phi}\approx 100$ TeV. In our case, standard masses in supergravity
scenarios,
$m_{\phi}\approx 1$ TeV, will be used.

On the other hand, in the scenario where the low reheating temperature
is obtained through an inflaton field, 
the result for the relic density is quite different
from the usual one with large $T_{RH}$. In fact, in certain cases, 
the usual relation   
$\Omega_{\tilde\chi_1^0} h^2 \propto  
1/\langle \sigma^{\mathrm{ann}}_{\tilde\chi_1^0} v \rangle$
is not even valid, and   
the relic abundance may well be proportional to
the annihilation cross section \cite{kolb}.
We will see below, however, that the relic abundance will not
increase when intermediate scales are considered.

\subsection{Inflation scenario}

%\vspace{0.2cm}
% {\bf 4.}
Let us consider for example the SUSY hybrid inflation
scenario studied in ref.~\cite{george}. 
There, the inflaton decay width can be computed with the result
\be 
\Gamma_{\phi}= 
\frac{1}{8 \pi }\left 
( \frac{m_{f}}{\langle \phi \rangle}\right )^2  m_{\phi}\ ,
\label{hybrid}
\ee
where 
$\langle \phi \rangle $ is the vacuum expectation value of the inflaton field, 
which is of the order of the unification scale,
$m_{\phi}$ is the inflaton mass, and
$m_{f}$ is the mass of the particle $f$ that the inflaton 
decay to (in this case a  right handed neutrino or sneutrino). 
Obviously,
$m_{f}$ should be smaller than the inflaton mass to allow 
for the decay 
$\phi \to f f$. 

Now, using eqs. (\ref{trh}) and (\ref{hybrid}) 
one obtains the following reheating temperature:
\bea 
T_{RH}= 1.7\times 10^{9}\ GeV  
\left(\frac{100\ GeV}{\langle \phi \rangle}\right)\ 
\left(\frac{m_f}{100\ GeV}\right)\ 
\left(\frac{m_{\phi}}
{100\ GeV}\right)^{1/2}\
\left(\frac{100}{g_*(T_{RH})} \right)^{1/4}\ .
\label{reheatingg}
\eea 
In ref.~\cite{linde} it
has been shown that an intermediate unification scale of the order of
$M_I\sim 10^{11}$ GeV is favoured by  inflation. 
Then, 
%in our example,  
recalling that the inflaton mass is constrained by 
$m_{\phi} \lsim M_I^2 / M_{Planck}$, we obtain
$m_{\phi}\sim 10^2$ GeV.  
From  Eq.(\ref{reheatingg}) we find that 
$T_{RH} \sim 1$ GeV, since now 
$\langle \phi \rangle\sim 10^{11}$ GeV.
This reheating temperature is lower than the 
typical freeze-out temperature
$T_F \simeq m_{\chi}/20$. 
Notice that in the standard GUT scenario discussed in
ref.~\cite{george}, one has $m_{\phi}\sim 10^{11}$
GeV, $\langle \phi \rangle\sim 10^{16}$ GeV, 
and therefore $T_{RH}\sim 10^{9}$ GeV.
%\cite{george}.

%==============================================

%In Ref. \cite{linde} it
%has been shown that an intermediate unification scale of the order of
%$M_I\sim 10^{11}$ GeV is favored by  inflation. 
%In this case, 
%recalling that the inflaton mass is constrained by 
%$m_{\phi} < M_I^2 / M_P$, one obtains
%($\langle \phi \rangle\sim 1.7\times 10^{16}$ GeV \cite{george})
% an inflaton mass of the order  $m_{\phi}\sim 10^2$ GeV.  For the 
%'standard GUT scenario' with $M_I \sim 10^{16}$ 
%one obtains however  $m_{\phi}\sim 10^{13}$ GeV . 
%From  Eq.(\ref{trh}) we find that 
%$T_{RH} \simeq \mathcal{O}(1)$ GeV;
%  a value much smaller than the values 
%quoted in the standard GUT scenario ($T_{RH}\sim 10^{11}$). 
% This reheating temperature is lower than the 
%typical freeze-out temperature $T_F$ which is given by 
%$T_F \simeq m_{\chi}/20$. 

%===============================================

As mentioned above,
a detailed  analysis of the relic density with a low reheating
temperature has been carried out in ref.~\cite{kolb} 
by Giudice, Kolb and Riotto. 
They study two possible non-relativistic cases,
depending 
on whether or not the dark-matter particles are
in chemical equilibrium. In the first one they are 
never in equilibrium, either before or after reheating.
In the second one the dark-matter particles
reach chemical
equilibrium, but then freeze out before the completion of the
reheat process. These scenarios not only lead to 
different qualitative and quantitative predictions
for the relic density, but also these predictions
are quite different from the
standard ones summarised in eq. (\ref{omega1}).

In the case of non--equilibrium production, 
the number density of neutralinos $n_{\tilde\chi_1^0}$ is much
smaller than $n_{\tilde\chi_1^0}^{eq}$, thus the relevant Boltzmann equations
% (\ref{boltzmann}) 
can be approximated and solved. 
One gets \cite{kolb}
\bea 
\Omega_{\tilde\chi_1^0} h^2 = 2.1 \times 10^4
\left(\frac{g}{2}\right)^2 \left(\frac{g_*(T_{RH})}
{10}\right)^{3/2} \left(\frac{10}{g_*(T_*)} \right)^3 \frac{(10^3
T_{RH})^7} {m_{\tilde\chi_1^0}^5} \left( \alpha_s + \frac{\alpha_p}{4}
\right)\ , 
\label{omega2} 
\eea 
where $g$ is the number of degrees of
freedom of the neutralino and $T_*$ is the temperature at which most of
the neutralino production takes place, it is given by 
$T_* \sim 4 m_{\tilde\chi_1^0} /15$. 
As we can see $\Omega_{\tilde\chi_1^0} h^2$ is proportional to
the annihilation cross section, instead of being inversely
proportional as in eq.~(\ref{omega1}). This raises the hope
that the relic abundance could be increased in scenarios
with intermediate scales where generically it is low. 
Unfortunately, the assumption
that $n_{\tilde\chi_1^0} << n_{\tilde\chi_1^0}^{eq}$ 
leads to a severe constraint on
the annihilation cross section \cite{kolb}. Namely $\alpha_s <
\bar{\alpha_s}$ and $\alpha_p < \bar{\alpha_p}$, where
$\bar{\alpha_s}$ and $\bar{\alpha_p}$ are of the order of 
$10^{-15}$ GeV$^{-2}$ and $10^{-14}$ GeV$^{-2}$, respectively,
for $T_{RH}\sim 1$ GeV. 
Since we are interested in large
cross sections, of the order of $10^{-7}$ GeV$^{-2}$, 
eq. (\ref{omega2}) cannot be applied. 

With a large annihilation
cross section ($\alpha_s > \bar{\alpha_s}$ or $\alpha_p >
\bar{\alpha_p}$), the neutralino reaches equilibrium before reheating as
discussed in ref.~\cite{kolb}, and its relic density is given by
\be 
\Omega_{\tilde\chi_1^0} h^2 = 2.3 \times
10^{-11} \frac{g_*^{1/2}(T_{RH})}{g_*(T_F)} \frac{T_{RH}^3 GeV^{-2}}
{m_{\tilde\chi_1^0}^3 \left(\alpha_s x_F^{-4} +4 \alpha_p x_F^{-5}/5\right)}\ ,
\label{omega3} 
\ee 
%
%where
%%
%\bea 
%x_F = \ln \left[ \frac{3}{ \sqrt{5} \pi^2} \frac{g~
%g_*^{1/2}(T_{RH})}{g_*(T_F)} \frac{M_P T_{RH}^2}{m_{\tilde\chi_1^0}} \left(
%\alpha_s x_F^{5/2} + \frac{5}{4} \alpha_p x_F^{3/2} \right) \right]\ .
%\eea 
%%
Now the relic density is again
inversely proportional to the annihilation cross section as in
eq. (\ref{omega1}). Moreover, it has a further suppression because of 
the low reheating temperature $T_{RH}\sim 1$ GeV, and as a consequence
%Moreover, it has a further suppression due to
%the very low reheating temperature effect.
we expect a result even worse than the one obtained in 
the standard computation discussed below eq. (\ref{omega1}).
Indeed, for a neutralino--nucleus cross section
of the order of $10^{-7}$
GeV$^{-2}$ we obtain $\Omega_{\tilde\chi_1^0} h^2 \approx 10^{-5}$.

\subsection{Modulus--decay scenario}
% {\bf 5.}
It has been
assumed in the above computation that
the relic abundance does not receive
any contribution from other sources. However, as shown
by Moroi and Randall \cite{moroi}, the production of 
neutralinos through moduli decay can modify those results.  

Let us recall first that moduli fields are present for 
example in string theory \footnote{See other 
examples of moduli fields, for instance in GUTs, in ref.~\cite{Lyth}.}.
Since moduli acquire masses through 
SUSY breaking effects, 
these masses, $m_{\phi}$, are expected to
be of the order of the gravitino mass, i.e.
$\cal{O}$($100 - 1000$ GeV). On the other hand,
their couplings with the MSSM matter are suppressed by a high energy scale. 
Thus one can parameterise the moduli decay width as 
\be
\Gamma_{\phi} = \frac{1}{2 \pi} \frac{m_{\phi}^3}{M_I^2}\ ,
\label{gphi}
\ee
where we denote with $M_I$ the effective suppression scale.
Since we are interested in the analysis of scenarios with intermediate
unification scales, we will consider the case of 
$M_I \approx 10^{11-14}$ GeV. An explicit example where this situation
arises is the case of type I string constructions.
There, twisted moduli are present with interactions suppressed by
the string scale. As discussed in ref.~\cite{benakli}
intermediate values for this scale can be obtained, 
and they have very interesting phenomenological 
implications \cite{benakli,Allanach,interm2,linde}.

\begin{figure}[t]
\centering
\epsfysize=70mm
\epsfxsize=100mm
%\epsfysize=100mm
%\epsfxsize=70mm
%\epsffile{a005.ps}
\epsffile{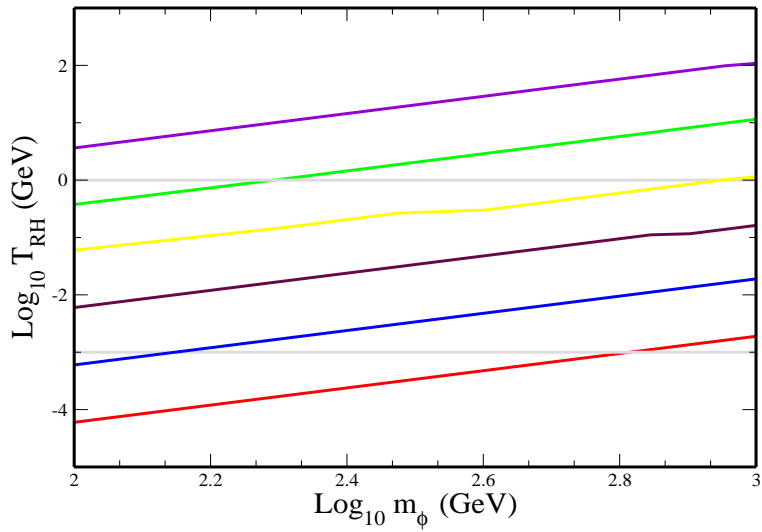}
\caption{ The reheating temperature $T_{RH}$ as a function of 
the  modulus mass $m_\phi$. The six curves correspond,
from top to bottom, to 
$M_I=10^{11},10^{12},10^{13},10^{14},10^{15},10^{16}$ GeV, respectively. 
The 
region bounded by the horizontal lines corresponds to a
reheating temperature larger than 1 MeV, a value close to the 
the nucleosynthesis limit, and smaller than 1 GeV, a value close to 
the freeze-out limit,
as explained in the text.}
\label{f1}
\end{figure}

Using eqs.~(\ref{trh}) and (\ref{gphi}) one obtains the
following reheating temperature:
\bea 
T_{RH}= 1\ MeV \times 
\left(\frac{6\times 10^{14}\ GeV}
{M_I}\right)\
\left(\frac{m_{\phi}}
{100\ GeV}\right)^{3/2}\
\left(\frac{10.75}{g_*(T_{RH})} \right)^{1/4}\ ,
\label{reheating}
\eea 
where note that $g_*=10.75$ for $T\sim$ $\cal{O}$($1-10$) MeV. 
This reheating temperature is 
shown as a function of the modulus mass in Fig.~\ref{f1} 
for different values of $M_I$.
The request that the modulus mass is larger than $\sim 100-500$ GeV in order 
to allow for kinematical 
decays into neutralinos of suitable mass $m_{\tilde\chi_1^0}\sim 50-200$ GeV,
limits in practice the reheating temperature 
to be above  $\sim 3$ GeV for the 
lowest scale on consideration $M_I=10^{11}$ GeV.
This has important consequences for the relic density computation.
As discussed in the introduction of this section,
we need 
a temperature smaller than the typical 
$T_F\simeq m_{\tilde\chi_1^0}/20\sim 3-10$ GeV, in
order to increase the relic neutralino density.
This implies that 
the scale $M_I\sim 10^{11}$ GeV is in the border of validity.
%Whereas values larger than this do work, e.g. 
On the other hand, for larger values we can obtain 
very easily interesting reheating temperatures. For example,
for $M_I=10^{12}$ GeV we have
$T_{RH}\gsim 0.3 $ GeV.
%, smaller values do not.
%we have to consider
%scales $M_I\gsim 10^{12}$ GeV, since
%$M_I= 10^{12}$ GeV corresponds to 
%$T_{RH}\gsim 0.3 $ GeV.
For the highest scale with an interesting
phenomenological value of the neutralino--nucleus cross section,
in the case of universality and moderate $\tan\beta$ \cite{interm1},
$M_I= 10^{14}$ GeV,
the lowest value of the reheating temperature corresponds to
$T_{RH}\sim 6 $ MeV. Larger values of the scale,
$M_I\gsim 10^{15}$ GeV, producing also a large
cross section, are possible in D-brane 
scenarios since non-universality in soft terms 
is generically present \cite{interm2}.
In this case
the constraint $T_{RH} \gsim$ 1 MeV from 
nucleosynthesis can be translated into a constraint 
on the modulus mass $m_{\phi}\gsim 140$ GeV.
% $M_I=10^{11}, 10^{12},10^{14},10^{16}$ GeV.

%It is worth noticing that we recover the usual result
%for $S$-- and $T$--moduli of string theory 
%using $M_I= M_{P}=2.4\times 10^{18}$ GeV.
%Then $T_{RH}$ above the nucleosynthesis limit implies
%a modulus mass above $25$ TeV.

When considering the 
decay of the modulus field producing neutralinos,
the evolution of the cosmological 
abundance of the latter becomes more complicated than in
the usual thermal-production case reviewed in Section~2.
Now one has to solve the coupled Boltzmann equations for the neutralino, 
the moduli field and the radiation \cite{moroi,chung}:
\bea 
\frac{d n_{\tilde\chi_1^0}}{d t} + 3 H n_{\tilde\chi_1^0}& =&
\bar{N}_{\tilde\chi_1^0} \Gamma_{\phi} n_{\phi} -\langle \sigma^{ann}_{\tilde\chi_1^0} 
v \rangle \left[ (n_{\tilde\chi_1^0})^2 -(n_{\tilde\chi_1^0}^{eq})^2 \right]\ ,
\label{moduli1}\\
\frac{d n_{\phi}}{d t} + 3 H n_{\phi} &=& - \Gamma_{\phi} n_{\phi}\ ,
\label{moduli2}\\
\frac{d \rho_{rad}}{d t} + 4 H \rho_{rad} &=& (m_{\phi} -
\bar{N}_{\tilde\chi_1^0} m_{\tilde\chi_1^0}) \Gamma_{\phi} n_{\phi} + 2 m_{\tilde\chi_1^0}
\langle \sigma^{ann}_{\tilde\chi_1^0} v \rangle \left[ (n_{\tilde\chi_1^0})^2
-(n_{\tilde\chi_1^0}^{eq})^2 \right]\ , 
\label{moduli3}
\eea 
%we follow the notation of Ref.~\cite{moroi}:  
%$m_\chi$ is the LSP mass, 
where $\bar{N}_{\tilde\chi_1^0}$ is the averaged number of neutralinos
produced in the decay of one modulus field. 
%The 
%number and energy densities of the modulus field are respectively 
%$n_\phi,\rho_\phi=m_\phi n_\phi$. 

Let us discuss qualitatively the solution following the arguments
used in ref.~\cite{moroi}.
For a $T_{RH}$ higher than $T_F$ the relic density will roughly
reproduce the usual result given by 
eq.~(\ref{thermal2}). However, for the interesting case for us
when $T_{RH}$ is lower than $T_F$, neutralinos produced
from modulus decay are never in chemical equilibrium, unlike the
thermal production case reviewed in Section~2.
As a consequence, its number density always decreases through
pair annihilation. When the annihilation rate 
$\langle\sigma^{ann}_{\tilde\chi_1^0} v \rangle n_{\tilde\chi_1^0}$
drops below the expansion
rate of the universe, $H$, 
the neutralino freezes out.
Then the relic density can be estimated as \cite{moroi}
\begin{eqnarray}
\Omega_{\tilde\chi_1^0}
%^{(ann)} 
h^2 & =& \frac{3\ m_{\tilde\chi_1^0}\ \Gamma_{\phi}}
{2\ (2 \pi^2/45)\ g_\star\ T_{RH}^3\  \langle 
\sigma_{\tilde\chi_1^0}^{ann} v  \rangle}\ 
\frac{h^2}{ \rho_c /s_0}\ .
\label{nlsp}
\end{eqnarray}
This result is valid when there is a large number of neutralinos
produced by the modulus decay. When the number is insufficient,
they do not annihilate and therefore all the neutralinos survive.
The result in this case is given by 
\begin{eqnarray}
\Omega_{\tilde\chi_1^0}
%^{(0)} 
h^2 & =& \frac{3\ \bar N_{\tilde\chi_1^0}\ m_\chi\ \Gamma_{\phi}^2\ M_P^2}
{(2 \pi^2/45)\ g_\star\ T_{RH}^3\ m_{\phi}}\ 
\frac{h^2}{ \rho_c /s_0}\ .
\label{nlsp2}
\end{eqnarray}
%
%Basically, eq.~(\ref{nlsp}) is valid for $N_{\tilde\chi_1^0}\sim 1$,
%whereas eq.~(\ref{nlsp2}) is valid for 
%$N_{\tilde\chi_1^0}\lsim 10^{-3}-10^{-4}$ \cite{moroi}.
The actual relic density is estimated  \cite{moroi} as the minimum of 
(\ref{nlsp}) and (\ref{nlsp2}).

\begin{figure}[h]
\begin{center}
%\centering
\begin{tabular}{c}
\epsfig{file= 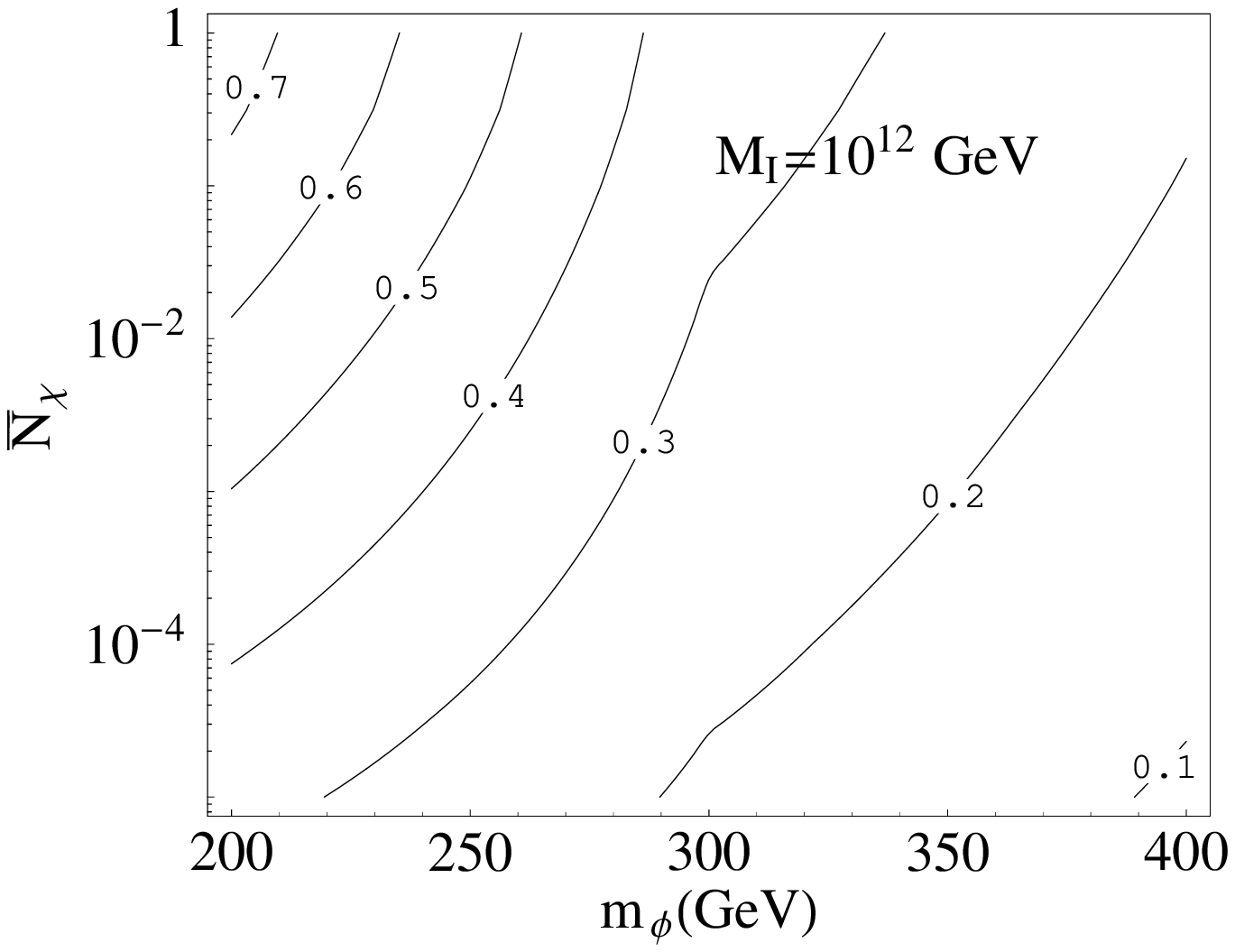, width=10cm, height=7.0cm}\\
% [-0.5cm]
\epsfig{file= 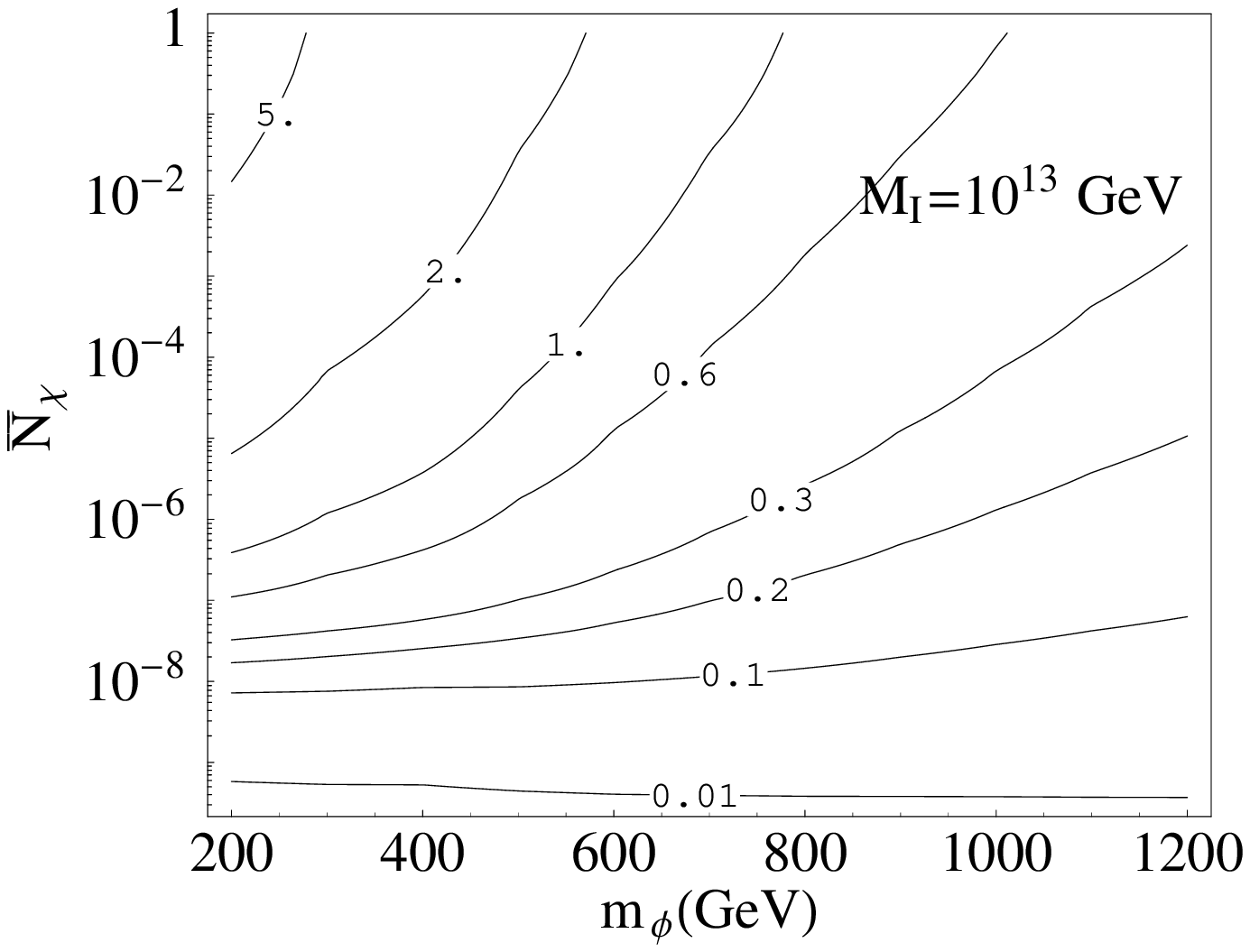, width=10cm, height=7.0cm}\\
% [-0.5cm]
\epsfig{file= 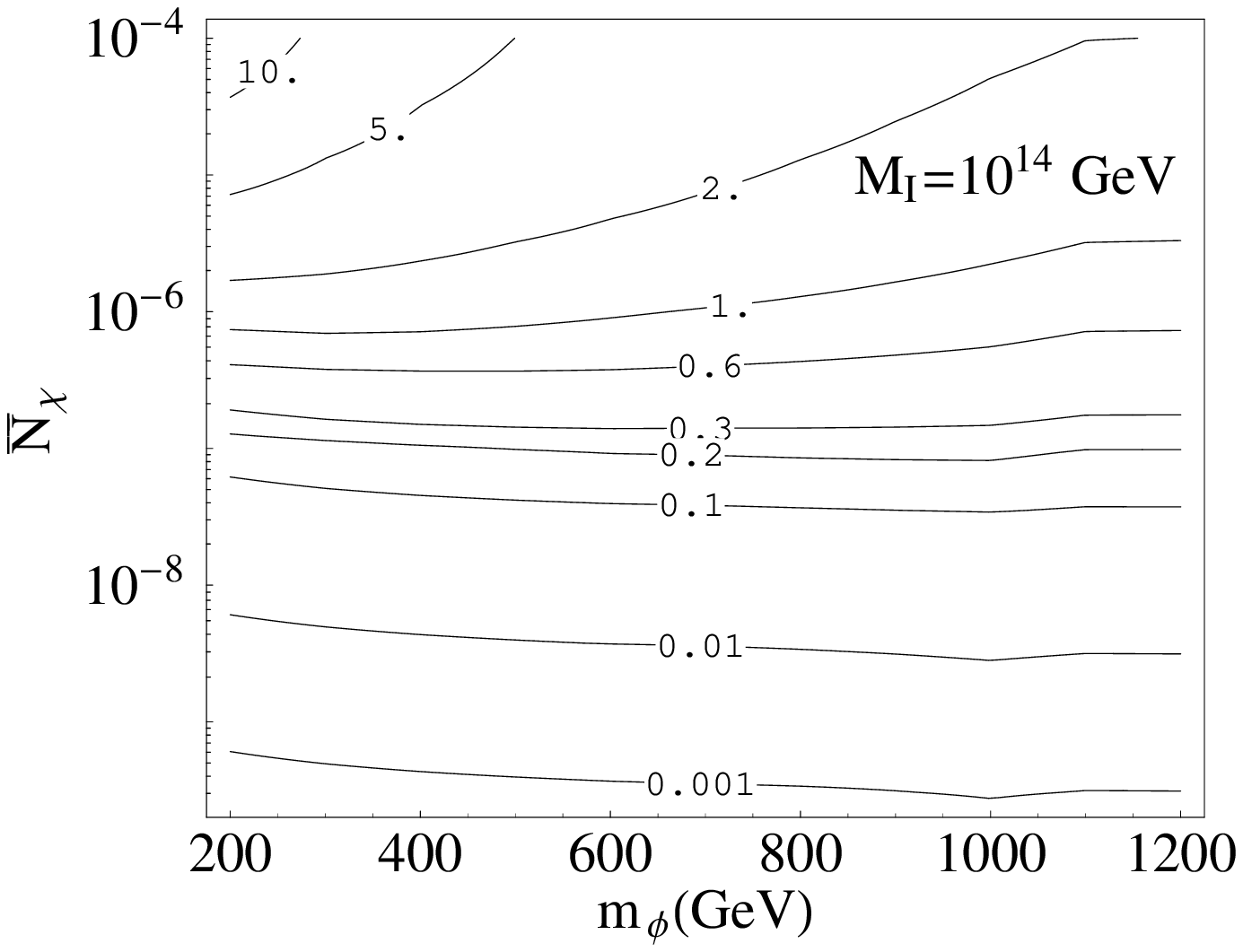, width=10cm, height=7.0cm}
\end{tabular}
\end{center}
%\centering
%\epsfysize=70mm
%\epsfxsize=100mm
%\epsffile{a12.ps}
%\epsffile{a13.ps}
%\epsffile{a14.ps}
\vspace{-0.7cm}
\caption{relic neutralino density $\Omega_{\tilde\chi_1^0} h^2$ 
contours as a function of $m_\phi$ 
and $\bar{N}_{\tilde\chi_1^0}$,
for $m_{\tilde\chi_1^0}=100$ GeV and 
several possible values of the intermediate scale 
$M_I=10^{12},10^{13},10^{14}$ GeV.
%The six curves correspond,
%from top to bottom, to 
%$M_I=10^{11},10^{12},10^{13},10^{14},10^{15},10^{16}$ GeV, respectively. 
%The 
}
\label{f2}
\end{figure}

%%
%\begin{eqnarray}
%\Omega_{\chi} h^2 \sim min \left(\Omega_{\chi}^{(ann)} h^2,\ \Omega_{\chi}^{(0)} h^2\right)\ ,
%\end{eqnarray}
%%
Now we can apply the above equations to our case with
intermediate scales.
Using eqs.~(\ref{gphi}) and (\ref{reheating}), 
%and the current value of
%the critical density, $\rho_c/s_0=3.6\times 10^{-9}$ GeV$\times h^2$, 
we can write expressions (\ref{nlsp}) and (\ref{nlsp2})
as
\bea 
\Omega_{\tilde\chi_1^0}
%^{(ann)} 
h^2 =  
\left(\frac{M_I}
{1.5\times 10^{20}\ GeV}\right)\
\left(\frac{1\ GeV^{-2}}
{\langle \sigma_{\tilde\chi_1^0}^{ann} v  \rangle}\right)
\left(\frac{100\ GeV}
{m_{\phi}}\right)^{3/2}\
\left(\frac{10.75}{g_*} \right)^{1/4}\
\left(\frac{m_{\tilde\chi_1^0}}
{100\ GeV}\right)
\ ,
\label{ann}
\eea 
\bea 
\Omega_{\tilde\chi_1^0}
%^{(0)} 
h^2 = 
\bar N_{\tilde\chi_1^0}
\left(\frac{1.2\times 10^{20}\ GeV}
{M_I}\right)\
%\left(\frac{1 GeV^{-2}}
% {\langle \sigma v  \rangle}}\right)
\left(\frac{m_{\phi}}
{100\ GeV}\right)^{1/2}\
\left(\frac{10.75}{g_*} \right)^{1/4}\
\left(\frac{m_{\tilde\chi_1^0}}
{100\ GeV}\right)
\ .
\label{(0)}
\eea 
From these equations 
%eq.~(\ref{ann}) 
we can see that even with a large annihilation cross section,
$\langle \sigma_{\tilde\chi_1^0}^{ann} v  \rangle\sim 10^{-8}$
GeV$^{-2}$, 
we are able to obtain the cosmologically interesting value  
$\Omega_{\tilde\chi_1^0}h^2\sim 1$.
For example for 
$M_I\sim 10^{13}$ GeV we obtain it 
%$\Omega_{\tilde\chi_1^0}\sim 0.1$
when $\bar N_{\tilde\chi_1^0}\sim 1$, using eq.~(\ref{ann}).
%For $M_I\sim 10^{14}$ GeV we obtain the above range
%with $\bar N_{\tilde\chi_1^0}\sim 10^{-7}$.
%In the case with insufficient neutralino production (\ref{(0)}) 
%we can also obtain those ranges but for very small values of
%$N_{\tilde\chi_1^0}$.
%For example for  $M_I\sim 10^{13}$ GeV we need
%$N_{\tilde\chi_1^0}\sim 10^{-7}-10^{-8}$.
In Fig.~\ref{f2} we show in more detail these results
solving numerically the Boltzmann eqs.~(\ref{moduli1})--(\ref{moduli3}),
for the large annihilation cross section introduced in eq.~(\ref{annihil2}).
There, the contours of constant relic neutralino density
$\Omega_{\tilde\chi_1^0} h^2$
as a function of $m_\phi$
%$m_{\tilde\chi_1^0}$ 
and $\bar{N}_{\tilde\chi_1^0}$
are shown, for fixed values of $M_I$ and $m_{\tilde\chi_1^0}$.
In particular, we consider the cases
$M_I=10^{12}, 10^{13}, 10^{14}$ GeV, with $m_{\tilde\chi_1^0}=100$ GeV.
The
corresponding reheating temperatures can be obtained from Fig.~1.
 %to  
%$T_{RH}=??, ??, ??$ GeV,
%respectively.
Note that whereas many values of 
$\bar N_{\tilde\chi_1^0}$ correspond to a satisfactory relic density
for 
$M_I=10^{12}-10^{13}$ GeV,
for the case $M_I=10^{14}$ GeV only a small range works.

Let us finally remark that the numerical value of 
$\bar N_{\tilde\chi_1^0}$ is in general model dependent.
This was discussed in the context of supergravity in
ref.~\cite{moroi}. In this particular case
both values
$\bar N_{\tilde\chi_1^0}\sim 1$ and 
$\bar N_{\tilde\chi_1^0}\sim 10^{-3}-10^{-4}$ are plausible,
depending on the 
characteristics of the supergravity theory under consideration.

%This expression is valid if the non-equilibrium production by the 
%modulus decay is not so small $\bar{N}_{LSP}>10^{-4}-10^{-5}$.
%The LSP relic density is roughly proportional to 
%$1/T_{RH} \langle \sigma v\rangle$, one recovers the standard result given
%by the expression \ref{radiation} when $T_{RH}>> T_F$.

%In Refs.\cite{interm1,interm2}, we obtained that for 
%$M_I\sim 10^{12}-10^{14}$ geV the 
%neutralino LSP acquires a sizeable higgsino component, as we commented before this is 
%essential for enhancing the LSP-nucleon cross section. 
%For a higgsino-like LSP the dominant annhilation chanel
% is the decay into a W boson  pair. 
%An upper bound on the cross section can be obtained \cite{moroi94}:
%\begin{eqnarray}
%\langle \sigma_{ann} v\rangle & <=& 
%\frac{\pi\alpha_2^2}{2}\frac{1}{m_\chi^2} \frac{(1-x_W)^{3/2}}
%{(2-x_W)^2},
%\end{eqnarray}
%where $\alpha_2$ is the $SU(2)$ constant and 
%$x_W=m_W^2/m_\chi^2$. For a LSP with 
%$m_\chi\sim 100$ GeV, the formula gives a maximum $\sim 2\times 10^{-8}$ 
%GeV$^{-2}$.
%In the case that the LSP have sizeable components of both gaugino 
%and higgsino one expects 
%smaller cross sections.

\clearpage

%\newpage

\section{Conclusions}

Current dark matter experiments are sensitive to 
a large neutralino-nucleus cross section, 
$\sigma_{\tilde\chi_1^0-N}\approx 10^{-7}$--$10^{-6}$ GeV$^{-2}$.
There are regions in the parameter space of SUSY scenarios with
an intermediate unification scale, where these cross sections can be obtained.
However, in these regions, the standard computation of the relic abundance
of neutralinos through thermal production in the early universe,
would imply a small relic abundance, 
$\Omega_{\tilde\chi_1^0} h^2 \lsim 0.01$.
%\vspace{0.2cm}
% {\bf 6.}

We have analysed some alternatives to solve this potential problem.
Let us recall that in the standard computation one is tacitly assuming
that the reheating temperature is much larger than the
freeze-out temperature, $T_F\sim 10$ GeV, and originated in an
inflationary
process. 
We have shown however that, when intermediate scales are present, reheating
temperatures as small as 
$\cal{O}$($1 - 1000$ MeV) are possible. 
Unfortunately, although the result for the relic abundance is modified,
still $\Omega_{\tilde\chi_1^0} h^2$ is too small.
On the other hand, when the above reheating temperatures
are associated to 
the decay of moduli fields producing out of equilibrium
neutralinos,
the relic abundance increases considerably with respect to 
the standard production. 
Thus the neutralino becomes a good dark matter
candidate with $0.1\lsim \Omega h^2 \lsim 0.3$, 
for intermediate scales $M_I\sim 10^{11}-10^{14}$ GeV, and 
moduli mass $m_\phi\sim 100-1000$ GeV.

%%%%We will consider the MSSM with
%%%%%% usual universality assumption. As mentioned above, the
%%%%%% neutralino--nucleon cross sections have been considered in this
%%%%%% model \cite{interm1} and it has been emphasized that in this case
%%%%%% one can get a large cross section at low $\tan \beta$. Here we fix
%%%%%% the parameters as in Ref. \cite{interm1}, namely we assume that
%%%%%% $\tan \beta = 5$ , the universal scalar mass $m_0$ is of order 150
%%%%%% GeV, and the trilinear coupling $\vert A \vert$ is taken to be
%%%%%% equal to the universal gaugino mass $M_{1/2}$. The values of $\mu$
%%%%%% and $B$ terms are determined from the electroweak breaking conditions.

\bigskip

\noindent {\bf Acknowledgements}

\noindent 
We would like to thank G. Lazarides for useful discussions. The 
work of S. Khalil was supported by the PPARC. 
The work of C. Mu\~noz was supported 
in part by the Spanish Ministerio de Ciencia y Tecnolog\'{\i}a
under contract FPA2000-0980, 
and
the European Union under contract HPRN-CT-2000-00148. 
The work of E. Torrente-Lujan was supported in part by  
the Spanish Ministerio de Ciencia y Tecnolog\'{\i}a
under contract FPA2000-0980, and by a MURST research grant.
%%%%%%%%%%%%%%%%%%%%%%%%%%%%%%%%%%%%%%%%

\end{document}